\documentclass{iau}
\usepackage{graphicx,subfigure,setspace,natbib}
\usepackage[usenames,dvipsnames]{color}
\usepackage{times}

\newcommand{\apj}{ApJ}           
\newcommand{\mnras}{MNRAS}       

\newcommand{\aap}{A\&A}
\newcommand{\araa}{ARA\&A}
\newcommand{\aj}{AJ}

\newcommand{\cplus}{\hbox{[C~\sc{ii}]}}
\newcommand{\hi}{\hbox{H~\sc{i}}}
\newcommand{\oi}{\hbox{[O~{\sc i}]}}
\newcommand{\lcii}{\hbox{L$_{\rm [C~{II}]}$}}
\newcommand{\lfir}{\hbox{L$_{\rm FIR}$}}

\title{A Herschel and CARMA view of CO and \cplus\ in Hickson Compact groups.}

\author[Alatalo, Appleton \& Lisenfeld (2014)]{Katherine Alatalo$^{1\star}$, Philip N.~Appleton$^1$ \and Ute Lisenfeld$^2$}

\affiliation{$^1$Infrared Processing and Analysis Center\\ California Institute of Technology, Pasadena, California 91125, USA \\
$\star$email: {\tt kalatalo@caltech.edu} \\
$^2$Departamento de F\'isica Te\'orica y del Cosmos, Universidad de Granada, Granada, Spain}

\pubyear{2014}
\volume{309}
\jname{Galaxies in 3D across the Universe}
\editors{B. L. Ziegler, F. Combes, H. Dannerbauer, M. Verdugo, eds.}

\begin{document}

\maketitle

\begin{abstract}
Understanding the evolution of galaxies from the starforming blue cloud to the quiescent red sequence has been revolutionized by observations taken with {\em Herschel} Space Observatory, and the onset of the era of sensitive millimeter interferometers, allowing astronomers to probe both cold dust as well as the cool interstellar medium in a large set of galaxies with unprecedented sensitivity.  Recent {\em Herschel} observations of of H$_2$-bright Hickson Compact Groups of galaxies (HCGs) has shown that \cplus\ may be boosted in diffuse shocked gas.  CARMA CO(1--0) observations of these \cplus-bright HCGs has shown that these turbulent systems also can show suppression of SF.  Here we present preliminary results from observations of HCGs with {\em Herschel} and CARMA, and their \cplus\ and CO(1--0) properties to discuss how shocks influence galaxy transitions and star formation.
\end{abstract}

\firstsection
\section{Introduction}
\begin{figure}[t!]
\centering
\includegraphics[height=0.75\textwidth,angle=90,clip,trim=1cm 0cm 2.5cm 0cm]{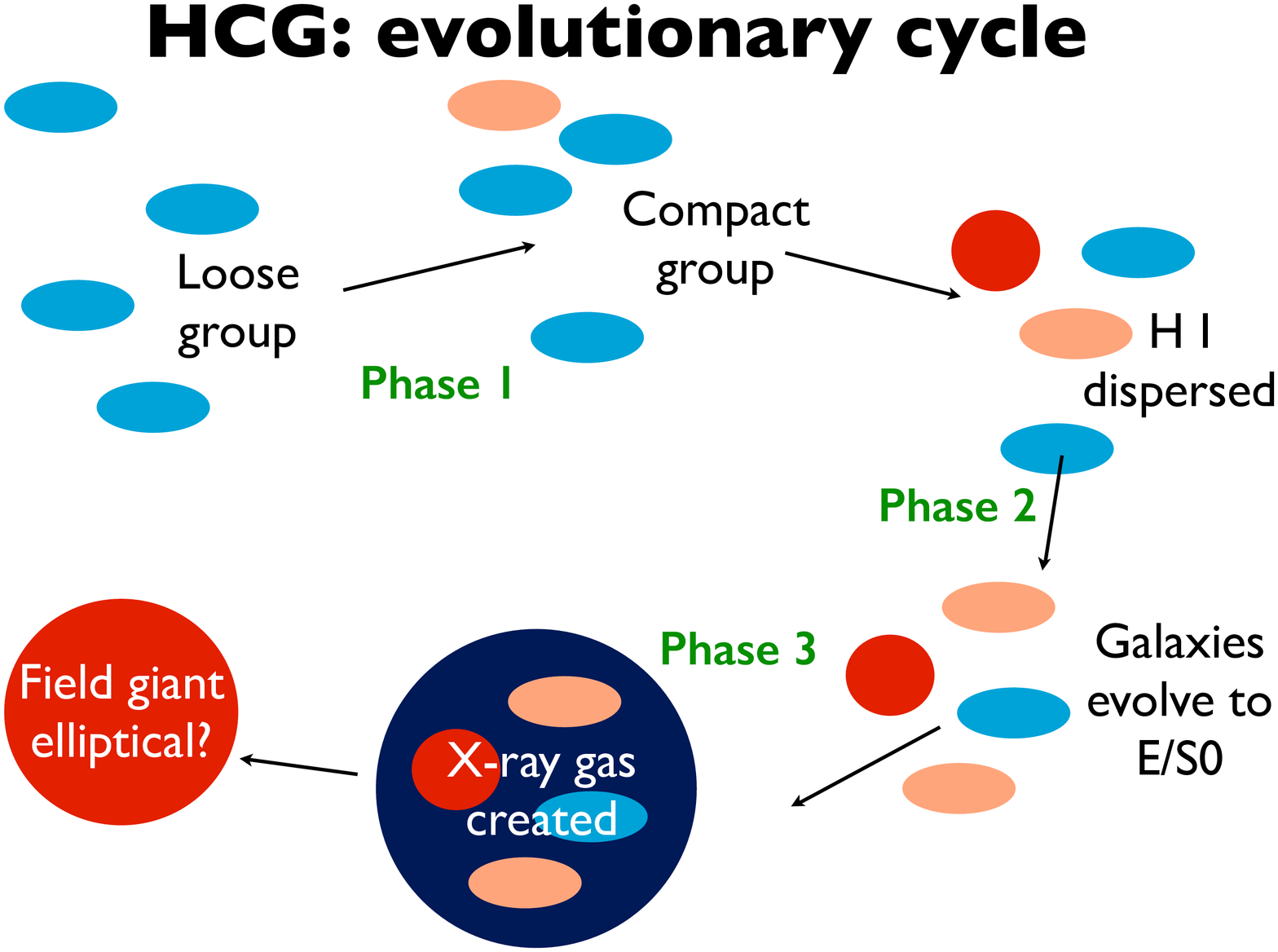} 
\caption{A schematic of the proposed HCG evolutionary picture.  The \hi\ observations taken by \citet{verdes-montenegro+01} and \citet{borthakur+10} support the \citet{hickson97} picture in which galaxies start in a loose group, move into the compavt group phase (during which the \hi\ in the individual galaxies are dispersed) until the gas is located within a common envelope.  During the evolution of the ISM, the galaxies within the group are also transitioning \citep{johnson+07, walker+10,walker+13,cluver+13}.}
\label{fig:evo_cyc}
\end{figure}

Compact groups are defined as ``small, relatively isolated systems of typically four or five galaxies in close proximity to one another'' \citep{hickson97}.  They tend to have a high fraction of early-type galaxies (E/S0), evidence of tidal interactions, are high density with low velocity dispersion \citep{hickson97} and are generally deficient in \hi\ \citep{verdes-montenegro+01,borthakur+10}.  Compact groups appear to go through an evolution (Fig. \ref{fig:evo_cyc}; \citealt{verdes-montenegro+01}) that can be traced based on its neutral gas depletion \citep{verdes-montenegro+01}, with galaxies the most advanced group stages no longer containing their own interstellar medium, and instead being surrounded by a common envelope \citep{rasmussen+08}.

\citet{johnson+07} documented a marked ``infrared gap'' in compact group galaxies, with very few galaxies observed between the star-forming cloud and the quiescent cloud.  \citet{cluver+13} showed that the ``gap'' galaxies in compact groups tended to have warm hydrogen emission (traced by the {\em Spitzer} Infrared Spectrograph) that was enhanced beyond the point that simple radiative processes could explained, deemed Molecular Hydrogen Emission Galaxies (MOHEGs; \citealt{ogle+06}).  \citet{cluver+13} were able to show that the enhanced H$_2$ emission was most easily explained via shocks.  With the onset of the Wide-field Infrared Survey Explorer (WISE), this infrared transition zone was shown to be more universal, with the bifurcation between early-type and late-type galaxies present in a large sample of galaxies \citep{alatalo+14}.  Understanding this transition takes both a broad survey approach, as well as carefully selected case studies.

Using the Photodetector Array Camera and Spectrometer (PACS; \citealt{pacs}) on {\em Herschel} \citep{herschel}, we observed ISM cooling lines, primarily \cplus\ and \oi\ of HCGs, and directly compare the ISM cooling to the molecular gas, traced via the Carbon Monoxide (CO) line, observed using the Combined Array for Research in Millimeter Astronomy (CARMA).

\section{Observations and Reduction}
\begin{figure*}
\includegraphics[width=\columnwidth]{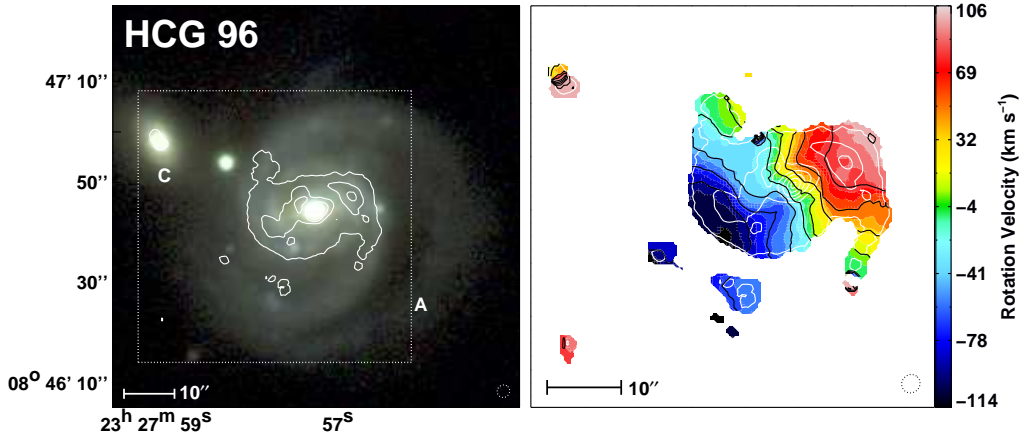} 
\caption{An example of the CO data from CARMA.  {\bf(Left):} The Sloan Digital Sky Survey {\em g, r, i} image underlies the moment0 map of HCG~96a and 96c (labeled on the diagram), which were both detected in the CARMA pointing. {\bf(Right):} The mean velocity map of HCG~96a and 96c.  In HCG~96a, the velocity structure traces the spirals, and in HCG~96c, the molecular gas is centralized, and regularly rotating.}
\label{fig:hcg96}
\end{figure*}

11 HCGs were observed with CARMA over the course of three semesters between 12 Mar 2013 -- 16 Jun 2014, and one, HCG~96 was taken from the archive, totaling 12 HCGs and 14 individual galaxies.  Raw data were reduced and calibrated identically to the ATLAS$^{\rm 3D}$ galaxies \citep{alatalo+13}.  Figure \ref{fig:hcg96} shows the integrated intensity and mean velocity map of HCG~96, as an example of the data quality that has been attained by the CARMA observations. CO in many of the HCGs have also been observed with single dish telescopes \citep{martinez-badenes+12,lisenfeld+14}.  For the most part, the fluxes from CARMA agree with the single dish fluxes, although in the cases of larger, more extended sources, CARMA recovers more flux than the single dish in cases where only a single pointing was observed.

{\em Herschel} PACS \cplus\ and \oi\ observations were taken of 11 HCGs, and the data were reduced using Herschel Interactive Processing Environment (HIPE) software package CIB13-3069, and data were analyzed identically to HCG~57, described in \citet{alatalo+14b}.  \cplus\ was detected in all 11 groups, and \oi\ was detected in 9.  Of the 11 PACS-imaged HCGs, 9 were detected in CO \citep{lisenfeld+14} and 8 were imaged using CARMA.

\section{Discussion}
\begin{figure*}
\vskip -3mm
\subfigure{\includegraphics[height=2.5in]{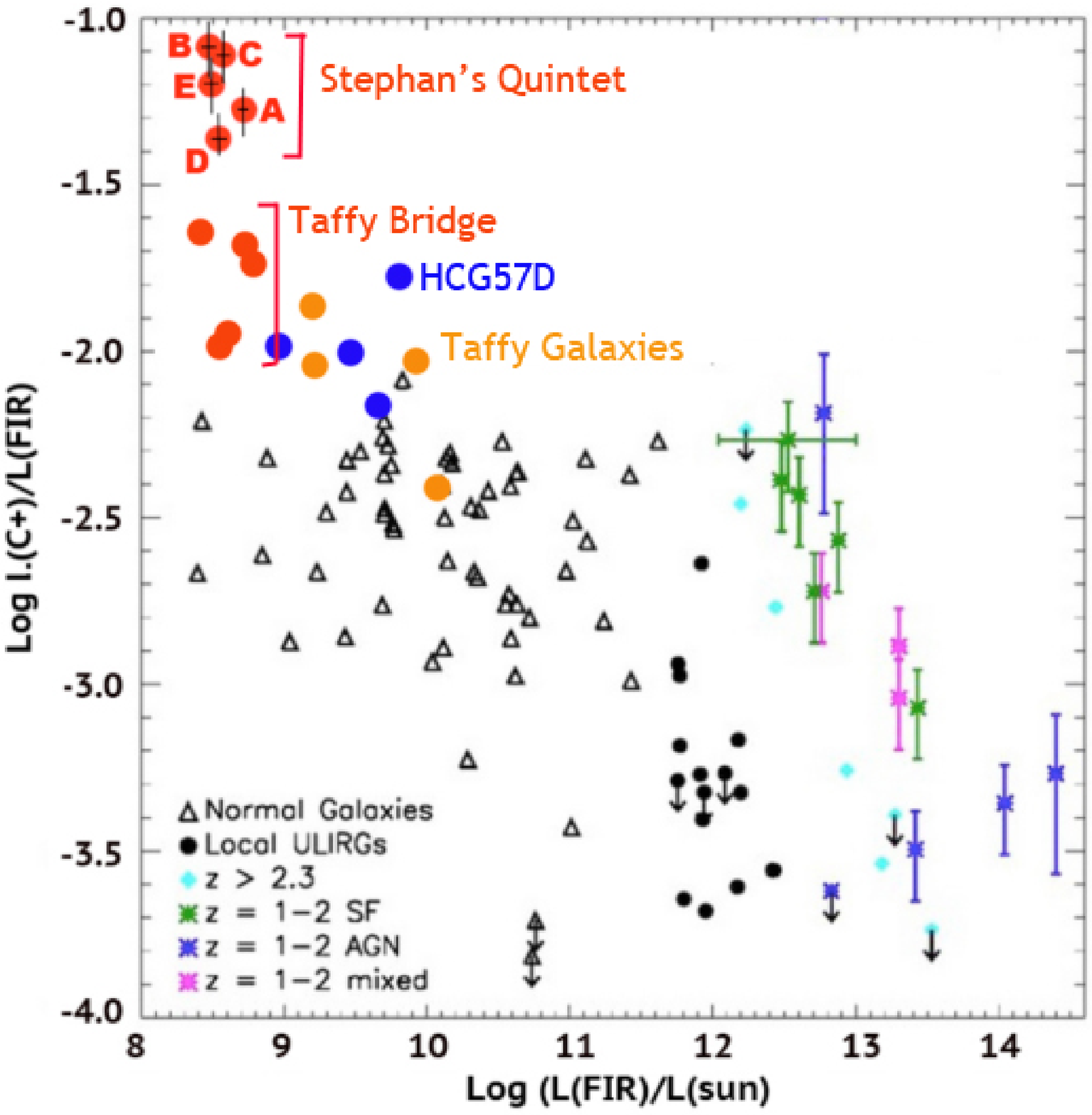}}
\subfigure{\includegraphics[height=2.5in]{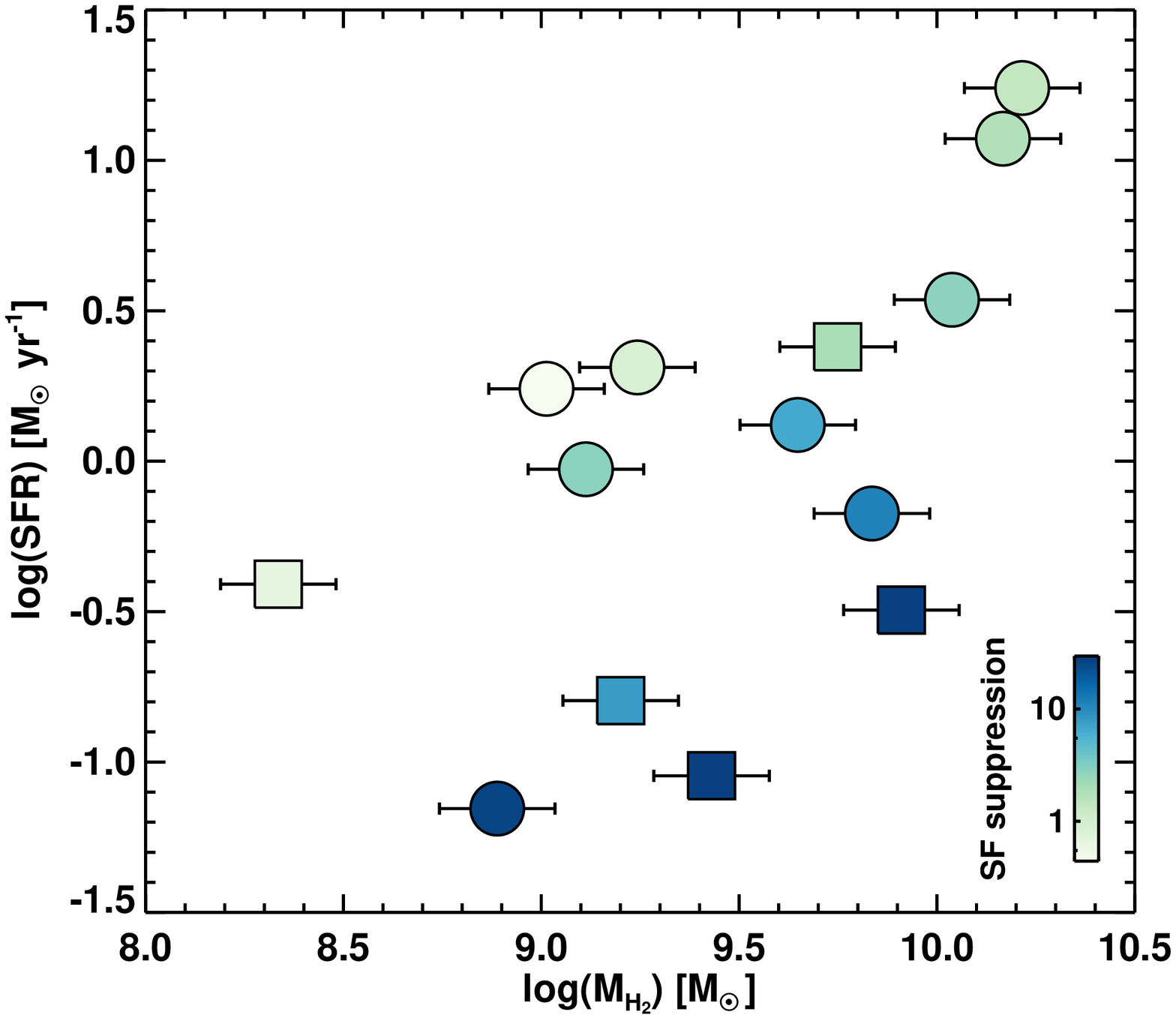}}
\caption{{\bf(Left):} The L(\cplus)/L$_{\rm FIR}$ vs L$_{\rm FIR}$ of starforming galaixes and ULIRGs \citep{malhotra+01} and high$-z$ sources \citep{stacey+10}.  There is an approximate L(\cplus)/L$_{\rm FIR}$ upper limit of 1\%, until we include sources known to have shocks, including the Taffy Galaxies \citep{peterson+12}, Stephan's Quintet \citep{appleton+13} and HCG~57 \citep{alatalo+14b}. {\bf(Right):} The total star formation rate calculated from \citet{bitsakis+14}, normalized to a Salpeter initial mass function \citep{salpeter55} compared to the molecular gas mass derived from the CO(1--0) data from CARMA (using the L$_{\rm CO}$--to--M$_{\rm H_2}$ conversion of $2.\times10^{20}$~cm$^{-2}$~(K~km~s$^{-1})^{-1}$; \citealt{bolatto+13}).  The colorization is based on the distance off of the relation derived in \citet{ken98}.}
\label{fig:c+sf}
\end{figure*}

In normal galaxies, \cplus\ can be a reliable tracer of SF \citep{delooze+11}, but this relation begins to break down as soon as one begins investigating the outliers.  Observations have shown that in local galaxies, \lcii/\lfir\ decreases in intense starburst environments (e.g. local ULIRGs; \citealt{malhotra+01,luhman+03}) and high-$z$ quasars, but the ``deficit'' may relate to the intensity of the UV radiation field.  \citet{diaz-santos+13} observed that \lcii/L$_{\rm FIR}$ decreases as a function of infrared luminosity in LIRGs, and thus in the most prolific star-formers, \cplus\ underestimates the true star formation rate. 
Figure \ref{fig:c+sf}a shows the inverse of the \citet{diaz-santos+13} result when one observes a selection of shocked galaxies (including Stephan's Quintet, Taffy and HCG~57).  In most ``normal'' galaxies, \lcii/\lfir\ rarely exceeds 1\%.  In these shock-dominated objects \citealt{peterson+12,cluver+13,appleton+13,alatalo+14b}, there are many more regions that exceed this 1\% ``ceiling''. These examples show that one must exercise caution when using \lcii\ to determine the SF in an object, but it also shows that \cplus\ is capable of pinpointing turbulence in systems.

Compact group galaxies whose ISM is turbulent, likely due to direct collisions that cause shocks, are also the galaxies that appear to contain the least efficient molecular gas.  This observation is supported by the fact that both Stephan's Quintet and HCG 57 show enhanced \cplus/L$_{\rm FIR}$ and signs of suppressed SF \citep{guillard+12,appleton+13,alatalo+14b}.  \citet{lisenfeld+14} studied the CO properties of MOHEGs and non-MOHEGs in HCGS, and, although several objects in their sample showed a suppressed SF, they did not find a statistical difference between the star formation efficiency of MOHEGs compared to  non-MOHEG sources. This might mean that excited H2 emission does not influence the star formation of an entire galaxy. 

Figure \ref{fig:c+sf}b compares the H$_2$ mass to the SF rate.  The colorization is based on the distance off of the Kennicutt-Schmidt relation ($\Sigma_{\rm SFR} = 2.5\times10^{-4} \Sigma_{\rm gas}^{1.4}$; \citealt{ken98}). Interestingly, this figure shows that many of the HCG galaxies  studied for this survey lay below the K-S relation, with several showing considerable ($>10$) suppression.  Two MOHEGs (including HCG~57a) represent the most suppressed objects. Given that galaxies in HCGs are much more likely to experience frequent interactions than field galaxies, they are ideal test beds for how turbulence can impact star formation. The fact that these systems are more likely to lie below the K-S relation than above is intriguing.  However, a larger number of objects and a more detailed study is neceesary to confirm this trend.  The \cplus\ imaging from Herschel will allow for a much more in-depth look at galaxies, determining areas of high turbulence within galaxies, and investigating whether those regions show signs of suppression. The recent study on HCG 57a, which used this method, seems to confirm the utility of this method. 

We are following up on these galaxies to determine where they lie in context, including whether they are found within the optical green valley or the infrared transition zone, to see if the suppression of star formation is a cause, or a symptom of this transition.  Further \cplus\ and CO(1--0) are needed of a larger sample of transitioning galaxies to fully understand this population, and in particular, how turbulence and shocks impact the supply of the cold molecular gas, as well as the how existing molecular gas forms stars.

\begin{spacing}{0.8}
{\small K.A. is supported by funding through Herschel, a European Space Agency Cornerstone Mission with significant participation by NASA, through an award issued by JPL/Caltech. U.L. acknowledges  support by the research projects AYA2011-24728 from the Spanish Ministerio de Ciencia y Educaci\'on and the Junta de Andaluc\'\i a (Spain) grants FQM108.  Support for CARMA construction was derived from the Gordon and Betty Moore Foundation, the Kenneth T. and Eileen L. Norris Foundation, the James S. McDonnell Foundation, the Associates of the California Institute of Technology, the University of Chicago, the states of California, Illinois, and Maryland, and the National Science Foundation. Ongoing CARMA development and operations are supported by the National Science Foundation under a cooperative agreement, and by the CARMA partner universities. {\em Herschel} is an ESA space observatory with science instruments provided by European-led Principal Investigator consortia and with important participation from NASA.  This work is based [in part] on observations made with the {\em Spitzer} Space Telescope, which is operated by the Jet Propulsion Laboratory, California Institute of Technology under a contract with NASA.}
\end{spacing}

\end{document}